\documentclass[aip,reprint,amsmath,amssymb]{revtex4-1}

\usepackage{graphicx}
\usepackage{dcolumn}
\usepackage{float}
\usepackage{bm}
\usepackage{amssymb}
\usepackage{lipsum}
\usepackage{color}

\newcommand{\be}{\begin{eqnarray}}
\newcommand{\ee}{\end{eqnarray}}
\newcommand{\bes}{\begin{eqnarray*}}
\newcommand{\ees}{\end{eqnarray*}}

\begin{document}

\title{Free-energy landscape of a polymer in the presence of two nanofluidic entropic traps}

\author{James M. Polson} 
\affiliation{ Department of Physics, University of Prince Edward Island, 
550 University Avenue, Charlottetown, Prince Edward Island, C1A 4P3, Canada }
\author{Matthew Kozma}
\affiliation{ Department of Physics, University of Prince Edward Island,
550 University Avenue, Charlottetown, Prince Edward Island, C1A 4P3, Canada }

\date{\today}

\begin{abstract}
Recently, nanofluidics experiments have been used to characterize the behavior of 
single DNA molecules confined to narrow slits etched with arrays of nanopits. Analysis
of the experimental data relies on analytical estimates of the underlying free-energy
landscape. In this study we use computer simulations to explicitly calculate the
free energy and test the approximations employed in such analytical models.
Specifically, Monte Carlo simulations were used to study a polymer confined to complex 
geometry consisting of a nanoslit with two square nanopits embedded in one of the surfaces. 
The two-dimensional Weighted Histogram Analysis Method (WHAM2D) is used to calculate
the free energy, {$F$}, as a function of the 
sum {($\lambda_1$)} and the 
difference {($\lambda_2$)} of {the length of}
the polymer contour contained in the two nanopits. We find the variation of the
free-energy function with respect to confinement dimensions to be comparable to
the analytical predictions that employ a simplistic theoretical model. {
However, there are some noteworthy quantitative discrepancies, particularly between
the predicted and observed variation of $F$ with respect to $\lambda_1$.
Our study provides a useful lesson on the limitations of using simplistic 
analytical expressions for polymer free-energy landscapes to interpret results
for experiments of DNA confined to a complex geometry and points to the value 
of carrying out accurate numerical calculations of the free energy instead.}

\end{abstract}

\maketitle

\section{Introduction}
\label{sec:intro}

Confinement of a polymer chain to a space with dimensions smaller than its radius
of gyration strongly affects its conformational and dynamical properties. Of particular 
note is the case of simple confinement geometries such as a slit or a long channel,
where a single length scale (i.e., slit height, channel width) characterizes the
degree of confinement. Over the past two decades, such systems have been extensively
examined experimentally using fluorescence microscopy of labeled DNA molecules.
As a consequence of this work, as well as that of studies of confined polymers 
using simulation and theoretical methods, 
the behaviour of these basic systems is now well understood.\cite{reisner2012dna, 
chen2016theory, dai2016polymer, frykholm2022dna, strychalski2012quantitative, 
lin2012effects, dorfman2014hydrodynamics, chen2014dynamics, tree2014odijk, 
leith2016free, cheong2018evidence, teng2021statistical, taylor2022confinement}

In addition to simple confinement geometries, there has also been increasing interest in 
characterizing the behaviour of polymers confined to more complex spaces featuring 
locally varying dimensionality and confinement scale. Examples include confinement 
of DNA in 2D arrays of spherical cavities connected by narrow 
pores,\cite{nykypanchuk2002brownian} as well as DNA stretched in a nanoslit 
between two microcavities or reservoirs.\cite{nykypanchuk2005single,
yeh2012entropy, kounovsky2013presentation, marie2013integrated, lameh2021electrokinetic}
Another example is a dual-nanopore device employing DNA ``flossing'' to 
exchange contour between two microcavities,  recently used to estimate the
contour distance between physical tags installed at sequence
motifs\cite{zhang2018single, liu2020flossing}
to provide electronic mapping of bacterial genome.\cite{rand2022electronic}

Another collection of studies have carried out
nanofluidics experiments to examine the configurational and dynamical 
behaviour of DNA confined to complex geometries composed of a nanoslit with parallel arrays
of nanogrooves\cite{mikkelsen2011pressure, vestergaard2016transition, smith2017photothermal}
and nanopits\cite{del2009pressure, klotz2012diffusion, klotz2015correlated, klotz2015measuring, 
klotz2016waves, kim2017giant} etched into one surface of the slit.
The pits and grooves act as entropic traps and generate a free-energy landscape that 
governs the equilibrium dynamics of the polymer. In the case of the nanopit geometry,
manipulation of the confinement dimensions (i.e., slit height and nanopit width, 
depth and separation) can create conditions wherein the polymer prefers to span two
or more pits.  From such a tuning of the free-energy landscape has emerged novel
phenomena such as diffusion resonance\cite{klotz2012diffusion} as well as 
collective wavelike excitations for excess concentration of DNA that span many
nanocavities.\cite{klotz2016waves} In addition, the dynamical behaviour of
DNA in such systems has been shown to provide a means to measure its effective
width.\cite{klotz2015measuring} 

Central to the analysis of the data from many DNA experiments employing such nanotopographies 
is an analytical estimate of the underlying free-energy landscape.
A typical approach is to construct composite theoretical models using established results 
for confinement in simple geometries. For example, a polymer with its contour spanning two
or more nanopits embedded in a slit surface can be divided into different sections, one for
each of the portions occupying the cavities and one for each connecting strand
lying in the slit between the cavities. Standard expressions for confinement free energy 
in cavities and slits can then be used to estimate the contribution from these sections.
A noteworthy example is the two-nanopit system of Ref.~\onlinecite{klotz2015correlated}.
This study examined the equilibrium dynamics by monitoring the fluctuations in the 
sum and the difference in the contour length of DNA trapped each of the two nanopits.
The fluctuations of these two quantities are governed by the two-dimensional 
free-energy function near its minimum. The analytical estimate of the free-energy function 
was used to explain the observed variation in the correlation times for the two modes with
respect to confinement dimensions.

\begin{figure}[!htb]
\begin{center}
\vspace*{0.2in}
\includegraphics[width=0.48\textwidth]{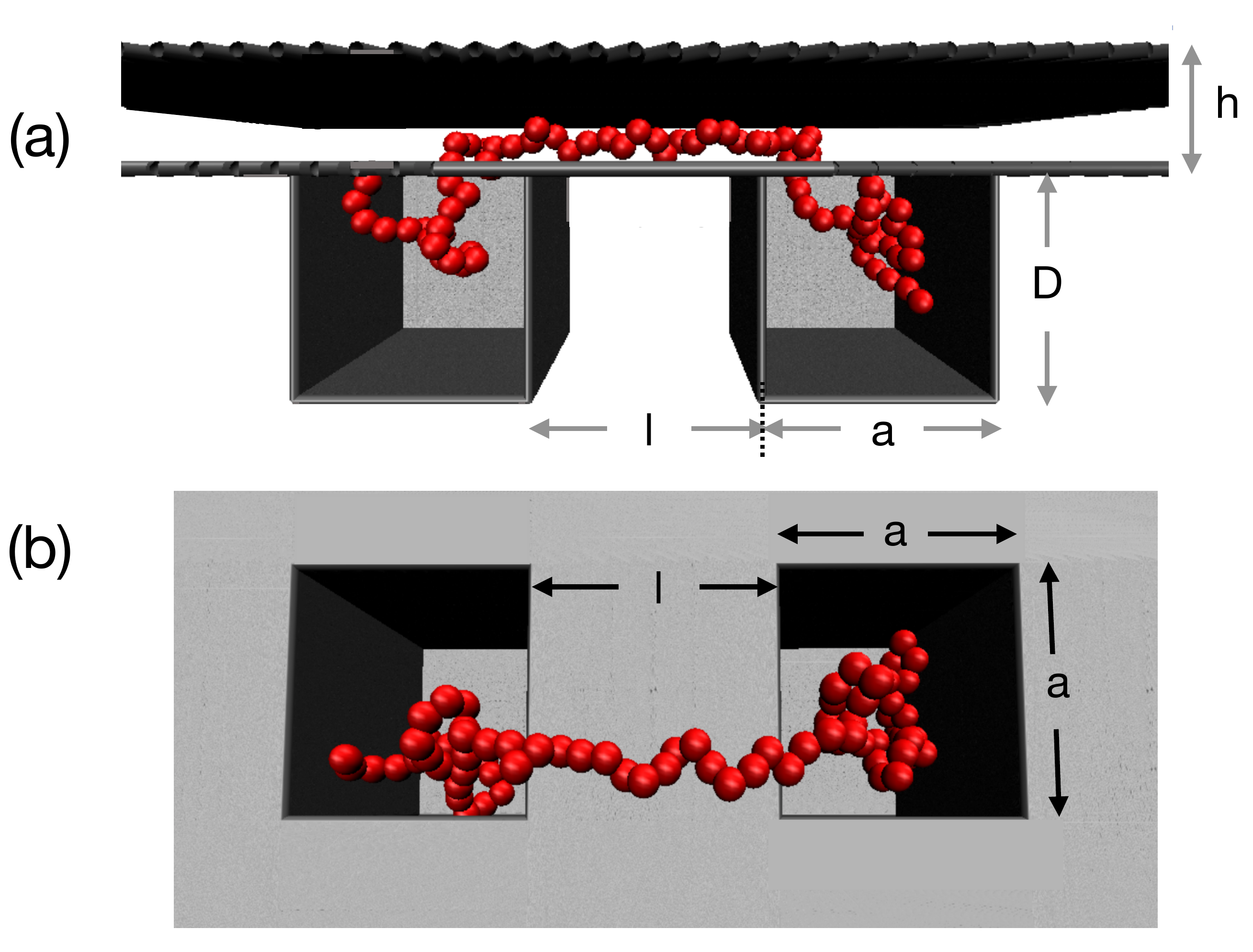}
\end{center}
\caption{Illustration of the model system employed in the study.
(a) Side view, showing the polymer spanning the two pits embedded in one
of the two surfaces of the slit. (b) Top view, with upper surface omitted,
for clarity.  }
\label{fig:illust}
\end{figure}

Analytical estimates for multi-dimensional free-energy functions for polymers 
confined to complex geometries such as that used in Ref.~\onlinecite{klotz2015correlated} 
are constructed using simplifying assumptions that may affect their accuracy and thus
their utility in the analysis of experimental data. For this reason, it would be
helpful to compare these analytical estimates with explicitly calculated free energies.
In the present study, we employ Monte Carlo (MC) simulations to calculate 
such free-energy functions for a simple model polymer confined to a two-nanopit 
geometry. {An illustration of the model system is shown in Fig.~\ref{fig:illust}.}
Lengthscale ratios for the polymer and confinement dimensions are comparable
to those employed in the experiments.
As in Ref.~\onlinecite{klotz2015correlated} we calculate the variation of 
$F$ with respect to the sum and the difference in the contour length in each of
the two pits. The two-dimensional free-energy function is obtained using the 2D Weighted
Histogram Analysis Method (WHAM2D).\cite{kumar1995multidimensional}
We find that the variation of the functions with system dimensions is comparable to
the trends observed in the experiments, though we do observe quantitative discrepancies
that originate from the breakdown in approximations employed in the theoretical model.

The remainder of the paper is organized as follows. Section~\ref{sec:model} describes
in detail the model system employed in the study.  Next, Sec.~\ref{sec:methods} gives 
an outline of the methodology used together with the relevant details of the simulations. 
In Section~\ref{sec:theory} we review that analytical model employed in 
Ref.~\onlinecite{klotz2015correlated} and tested in this work.
Section~\ref{sec:results} presents and discusses the simulation results. 
Finally, Sec.~\ref{sec:conclusions} summarizes the main conclusions of this work.

\section{Model}
\label{sec:model}

The system is comprised of a single polymer chain confined to a slit bounded by
two infinite parallel planar surfaces separated by a distance $h$. Embedded in
one of the surfaces are two identical pits. The pits each have a square cross section 
in the lateral directions with a side length of $a$. In addition, they have a constant 
depth of $D$, as measured in the longitudinal direction from the surface into 
which they are embedded. The nearest edges of the pits are parallel to each other and 
are separated by a distance $l$. {The various dimensions associated with
the confining geometry are illustrated in Fig.~\ref{fig:illust}.}

The linear polymer is modeled as a semiflexible chain of $N$ hard spheres of diameter $w$, 
which is thus the width of the polymer chain.  The bending rigidity of the polymer
is modeled using a bending potential with the form, $u_{\rm bend}(\theta)=
\kappa(1 - \cos\theta)$. The angle $\theta$ is defined for a consecutive
triplet of monomers centered at monomer $i$ such that
$\cos\theta_{i}=\hat{u}_{i}\cdot\hat{u}_{i+1}$,
where $\hat{u}_{i}$ is the unit vector pointing from monomer $i-1$ to monomer $i$.
The bending constant $\kappa$ determines the overall stiffness of the polymer and is
related to the persistence length $P$ by\cite{micheletti2011polymers}
$\exp(-\langle l_{\rm bond} \rangle/P) = \coth(\kappa/k_{\rm B}T) - k_{\rm B}T/\kappa$,
where $\langle l_{\rm bond}\rangle$ is the mean bond length.
For our model, the bond length is fixed to $l_{\rm bond}=w$.
Note that for sufficiently large $\kappa/k_{\rm B}T\gg 1$ this implies $P/w\approx
\kappa/k_{\rm B}T$. In addition to the polymer bending energy, there are two additional
contributions to the total energy that are both athermal in character: the 
interactions between non-bonded monomers and the monomer-wall interaction. In
each case, the energy is infinite in the case of overlap and zero otherwise.
System configurations with infinite energy, of course, have zero probability and so 
are forbidden.

Simulation employed polymer chains of length $N=1600$ monomers. Typically, the 
bending rigidity was fixed to $\kappa=5k_{\rm B}T$, though in some calculations this value
was varied. In addition, most calculations employed confining geometry parameter
values of $h=10w$, $D=10w$, $a=60w$, and $l=60w$. To map the model system
onto the experimental system by choosing an effective DNA width of $w=10$~nm,
this corresponds to a contour length of $L$=16~$\mu$m and a persistence length
of $P=50$~nm, as well as confining geometry parameter values of $h=100$~nm, $D=100$~nm,
(i.e., $d=h+D=$200~nm), $a=600$~nm, and $l=600$~nm. These values are 
comparable to the values used in the experiments of Refs.~\onlinecite{klotz2015correlated} 
and \onlinecite{klotz2015measuring}. Simulations were carried out in which each
of these parameters were independently varied with the others each fixed to 
these ``standard'' values.

\section{Methods}
\label{sec:methods}

Computer simulations were used to calculate a collection of two-dimensional free-energy
functions. In particular, the configurational free energy $F$ is measured as a function
of the two quantities $\lambda_1\equiv N_1+N_2$ and $\lambda_2\equiv N_1-N_2$,
where $N_1$ and $N_2$ are the numbers of spherical monomers whose centers lie inside
the pits, arbitrarily labeled ``1'' and ``2''. A monomer center is defined to lie inside 
the pit if it lies within the square cross section of the pit in the lateral plane, independent
of its position in the vertical direction.  All simulations are carried out using 
the standard Metropolis Monte Carlo method.{\cite{frenkel2002understanding, allenbook}}
We employ a combination of crankshaft and reptation trial moves. In each case, trial
moves are rejected if a proposed monomer position overlaps with a surface of the
confining geometry. In the absence of such overlap crankshaft moves are accepted 
with a probability given by the usual Metropolis criterion, where the energy difference
between old and new configurations is determined solely from the internal bending
energy of the chain. In each simulation the system was equilibrated for a period
of $5\times 10^6$ MC cycles, following which quantities were sampled during a 
production run of $1.1\times 10^8$ MC cycles. One MC cycle corresponds to $N$
attempted crankshaft moves and 10 attempted reptation moves.

In order to improve computational efficiency, we employ the method of cell lists.
Since the system is unbounded in the $x-y$ direction, we employ periodic boundaries
in these two dimensions using a square simulation box of side length 240$w$. 
Extensive testing showed that this box size was sufficiently large to minimize 
artificial self-interaction through the periodic boundaries. The cell list method
improved the computational efficiency by roughly a factor of two.

In order to calculate the free-energy function, we also use the 2D Weighted Histogram 
Analysis Method (WHAM2D).\cite{kumar1995multidimensional} The procedure requires
that a set of individual simulations be carried out, each subject to a unique biasing
potential. In each simulation, histories of the values of $\lambda_1$ and $\lambda_2$ 
are calculated,
and the WHAM2D method constructs an unbiased probability distribution using these
histories as input. The method is useful for overcoming problems of 
poor statistics, particularly in low probability regions. Each calculation employs
a quadratic biasing potential of the form
\begin{eqnarray} 
U_{\rm b}={\textstyle{\frac{1}{2}}}k_1 (\lambda_1-\lambda_{1}^{\rm (c)})^2
+ {\textstyle{\frac{1}{2}}}k_2 (\lambda_2-\lambda_{2}^{\rm (c)})^2.
\end{eqnarray}
The coefficients of the potential were chosen to be $k_1=k_2=0.005k_{\rm B}T/w^2$. In addition,
the biasing potential centers were chosen to be spaced by $\Delta \lambda_1^{\rm (c)}=10w$
over the interval $\lambda_1\in(0,N)$ and $\Delta \lambda_2^{\rm (c)}=20w$
over a range of $\lambda_2\in[-\lambda_1,\lambda_1]$. The latter range 
follows from the obvious restriction that no pit can contain a greater number of 
monomers than the total number in both pits. These values of the biasing potential
parameters ensure that there is sufficient overlap in the probability distributions
between neighboring ``potential centers'', as determined by the locations
of $\lambda_1^{\rm (c)}$ and $\lambda_2^{\rm (c)}$. {Further details
about the implementation of the WHAM2D method are presented in Appendix~\ref{app:a}.}
In most cases, a calculation
to produce a single 2D free energy function required 6801 potential centers,
and thus the same number of simulations. As each simulation required approximately
50--60 CPU-hours, each free energy function required up to 47 CPU-years.

In the results presented below, distances are measured in units of $w$ and energy
is measured in units of $k_{\rm B}T$.

\section{Theory}
\label{sec:theory}

In this section we employ simple theoretical arguments to predict the functional
form of $F(\lambda_1,\lambda_2)$ following the approach in Refs.~\onlinecite{klotz2015correlated} 
and \onlinecite{klotz2015measuring}. 

Consider a ideal Gaussian chain of $N_{\rm K}$ Kuhn segments of length $b$ confined to 
a rectangular box with dimensions $L_x$, $L_y$ and $L_z$. As noted in 
Ref.~\onlinecite{Doi_book_1986}, the conformational contribution to the configurational 
partition function is given by:
\be
\frac{F_{\rm id,box}}{k_{\rm B}T} & = &   
\frac{\pi^2 N_{\rm K}b^2}{6}  \left(\frac{1}{L_x^2}+\frac{1}{L_y^2}+\frac{1}{L_z^2}\right).
\label{eq:Fall}
\ee
%
%
Thus, the confinement free energy of the polymer confined to a box of dimensions $L_x=L_y=a$ 
and $L_z=d$ is given by
\be
\frac{F_{\rm id,box}}{k_{\rm B}T} & = & A_{\rm box} L,
\label{eq:Fidbox}
\ee
where
\be
A_{\rm box} & = & \frac{\pi^2 P}{3} \left(\frac{1}{d^2}+\frac{2}{a^2}\right),
\label{eq:deff1box}
\ee
and where $L=N_{\rm K}b$ is the contour length and $P=b/2$ is the persistence length.
Although derived for a Gaussian chain, 
this expression is expected to be valid for wormlike chains in the limit where the confinement 
dimensions are large, i.e. $P\ll d$ and $P\ll a$. In the case where this condition is
only marginally satisfied, a numerical calculation is required for a more accurate measure
of the free energy. In Ref.~\onlinecite{klotz2015measuring}, Klotz {\it et al.} performed
such a calculation for a wormlike chain confined a pillbox cavity (see Fig.~2(b) of that
article).  For a real chain, excluded-volume interactions
can be incorporated at the second-virial level,
${F}_{\rm EV}/k_{\rm B}T = BL^2$,
where
\be
B(d,a,w) \equiv \frac{\pi}{4}\frac{w}{V_{\rm c}},
\label{eq:deff2}
\ee
and where $V_{\rm c}=da^2$ is the volume of the rectangular cavity. The sum of the two terms, 
$F_{\rm box} \equiv F_{\rm id,box}+F_{\rm EV}$ is an approximation for the difference
in the conformational free energy of a confined real chain and an unconfined ideal chain.
This can be written:
\begin{eqnarray}
F_{\rm box} = A_{\rm box} L + B L^2
\label{eq:Fbox}
\end{eqnarray}
where $A_{\rm box}$ and $B$ are given by Eqs.~(\ref{eq:deff1box}) and 
(\ref{eq:deff2}), respectively.

Now consider the case of an ideal polymer confined to a slit of height $h$. 
Retaining only the last term of Eq.~(\ref{eq:Fall}) and choosing $L_x = L_y\rightarrow\infty$ 
and $L_z=h$, the confinement free energy is given by
${F_{\rm id,slit}}/{k_{\rm B}T} = \pi^2 L P/3h^2$.
This expression is expected to be valid for a wormlike chain in the limit where $P\ll h$.
In the opposite limit where $h\ll P$, the chain is in the Odijk regime. Here, the exact 
expression for the free energy is given by\cite{burkhardt1997free}
${F_{\rm id,slit}}/k_{\rm B}T = 1.104L(Ph^2)^{-1/3}$.
The intermediate regime between these two limiting cases is well approximated by an
interpolation obtained by Chen and Sullivan,\cite{chen2006free} who solved the worm-like
chain diffusion equation in a slit geometry with hard-wall boundary conditions imposed at 
the slit surfaces. Their result is well approximated by the analytical expression
\be
\frac{F_{\rm id,slit}}{k_{\rm B}T} = A_{\rm slit} L,
\label{eq:FCS}
\ee
where
\be
\hspace*{-0.25in}
A_{\rm slit} = \frac{\pi^2}{3}\frac{P}{h^2}\left(5.14\left(\frac{P}{h}\right)^2
                      +  1.98\left(\frac{P}{h}\right) + 1\right)^{-2/3}.
\label{eq:deff2CS}
\ee
For sufficiently small $h$, excluded-volume interactions are negligible. Thus, Eq.~(\ref{eq:FCS}) 
is approximately the difference in the conformational free energy of a slit-confined real
chain and an unconfined ideal chain.

A polymer whose contour spans two nanopits can be viewed as two contour clusters, each centered
on a single pit and connected with a linking strand that contributes a spring free energy, 
$F_{\rm spr}$.  As in Refs.~\onlinecite{klotz2015correlated} and \onlinecite{klotz2015measuring}, 
we approximate this free energy with 
a modified form of the Marko-Siggia force equation that accounts for the effects of vertical
confinement in the nanoslit and was obtained from a parameterization of MC simulation 
data:\cite{dehaan2015force}
\begin{eqnarray}
\frac{F_{\rm spr}}{k_{\rm B}T} = \frac{l^2}{2PL_{\rm s}} \frac{D_{\rm eff}-1}{4(L_{\rm s}-l)} 
                                 [2L_{\rm s} D_{\rm eff} - l(D_{\rm eff}+1)]
\label{eq:Fspr}
\end{eqnarray}
Here, $L_{\rm s}$ is the contour length of the connecting strand, and the quantity $l$ is the 
distance between the ends of the connecting strand, which for simplicity we choose here to be the 
spacing between the parallel edges of the nanopits.  The effective dimensionality, $D_{\rm eff}$, 
given by 
\begin{eqnarray}
D_{\rm eff}(h) = 1 + \frac{2}{2-\exp(-0.882(P/h)^{1.441})},
\end{eqnarray}
interpolates between dimensions of 2 and 3 for small and large slit height, $h$, respectively.
As in the original Marko-Siggia relation, this free energy only becomes appreciably larger
than $k_{\rm B}T$ when $L_{\rm s}$ is not significantly larger than $l$.

Now consider a polymer chain of contour length $L$ that spans two nanopits of depth $d$, side 
length $a$, and spacing $l$.  
The total conformational confinement free energy of the polymer can be approximated
\begin{eqnarray}
F_{\rm conf} = F_{\rm cav} - F_{\rm id,slit} + F_{\rm spr},
\label{eq:Fconf}
\end{eqnarray}
where $F_{\rm id,slit}$ is given by Eq.~(\ref{eq:FCS}) and $F_{\rm spr}$ is given by
Eq.~(\ref{eq:Fspr}). In addition, $F_{\rm cav}$ is the free energy increase associated
with confinement of parts of the polymer to the nanopits and is given by
\begin{eqnarray}
F_{\rm cav}(L_1,L_2) = F_{\rm box}(L_1) + F_{\rm box}(L_2),
\label{eq:Fcav}
\end{eqnarray}
where $F_{\rm box}$ is given by Eq.~(\ref{eq:Fbox}). The first term accounts for confinement
of one subchain of length $L_1$ in one nanopit and another subchain of length $L_2$ in
the other nanopit. Combining Eqs.~(\ref{eq:Fbox}), (\ref{eq:Fconf}) and (\ref{eq:Fcav}), it 
can be shown that
\be
\frac{F_{\rm conf}}{k_{\rm B}T} = -A (L_1+L_2) + B (L_1^2 + L_2^2) + F_{\rm spr}(L_{\rm s})
\label{eq:Fconf2}
\ee 
where 
\be
A \equiv A_{\rm slit} - A_{\rm box}. 
\label{eq:Adef}
\ee

Defining $\lambda_1\equiv L_1+L_2$ and $\lambda_2\equiv L_1-L_2$, i.e., the sum of and the
difference in the contour lengths in each of the nanopits, Eq.~(\ref{eq:Fconf2}) can be
written:
\be
\frac{F_{\rm conf}(\lambda_1,\lambda_2)}{k_{\rm B}T} = -A \lambda_1 
         + {\textstyle\frac{1}{2}} B \lambda_1^2 + {\textstyle\frac{1}{2}} B\lambda_2^2
         + F_{\rm spr}(L-\lambda_1), \nonumber \\
&&
\label{eq:Fconf3}
\ee
where we note the argument $L_{\rm s}= L-\lambda_1$ in the last term.
Equation~(\ref{eq:Fconf3}) can be written
\be
\frac{F_{\rm conf}(\lambda_1,\lambda_2)}{k_{\rm B}T} & = & 
\frac{\kappa_{\rm s}}{2} (\delta\lambda_1)^2 + \frac{\kappa_{\rm a}}{2} (\lambda_2)^2 
+ {\cal O}(\delta\lambda_1^3).
\label{eq:Fsumdiff}
\ee
Here, $\delta\lambda_1\equiv \lambda_1-\lambda_1^{\rm (min)}$ and 
$(\lambda_1,\lambda_2)=(\lambda_{1}^{\rm (min)},0)$ is the location of the free energy 
minimum. The quantities  $\kappa_{\rm s}$ and  $\kappa_{\rm a}$ are effective
spring constants that govern the fluctuations in the sum and difference of the contour
lengths in the two pits, which are symmetric and asymmetric modes, respectively. 
Note that $\kappa_{\rm a}=B$. The other coefficient, $\kappa_{\rm s}$, does not 
have a cromulent closed-form expression in terms of the system parameters.
However, in the limit of sufficiently small $\lambda_1$, the spring
free energy term in Eq.~(\ref{eq:Fsumdiff}) is negligible. In this case,
Eq.~(\ref{eq:Fconf3}) can be rewritten:
\be
F/k_{\rm B}T \approx {\textstyle\frac{1}{2}} B \left(\lambda_1-A/B\right)^2 
+ {\textstyle\frac{1}{2}} B(\lambda_2)^2 + {\rm const}
\label{eq:Fsea}
\ee
Comparing Eqs.~(\ref{eq:Fsea}) and (\ref{eq:Fsumdiff}), it follows that
$\kappa_{\rm s} = \kappa_{\rm a}$ and $\lambda_1^{\rm (min)}=A/B$.

\begin{figure*}[!htp]
\begin{center}
\vspace*{0.2in}
\includegraphics[width=0.95\textwidth]{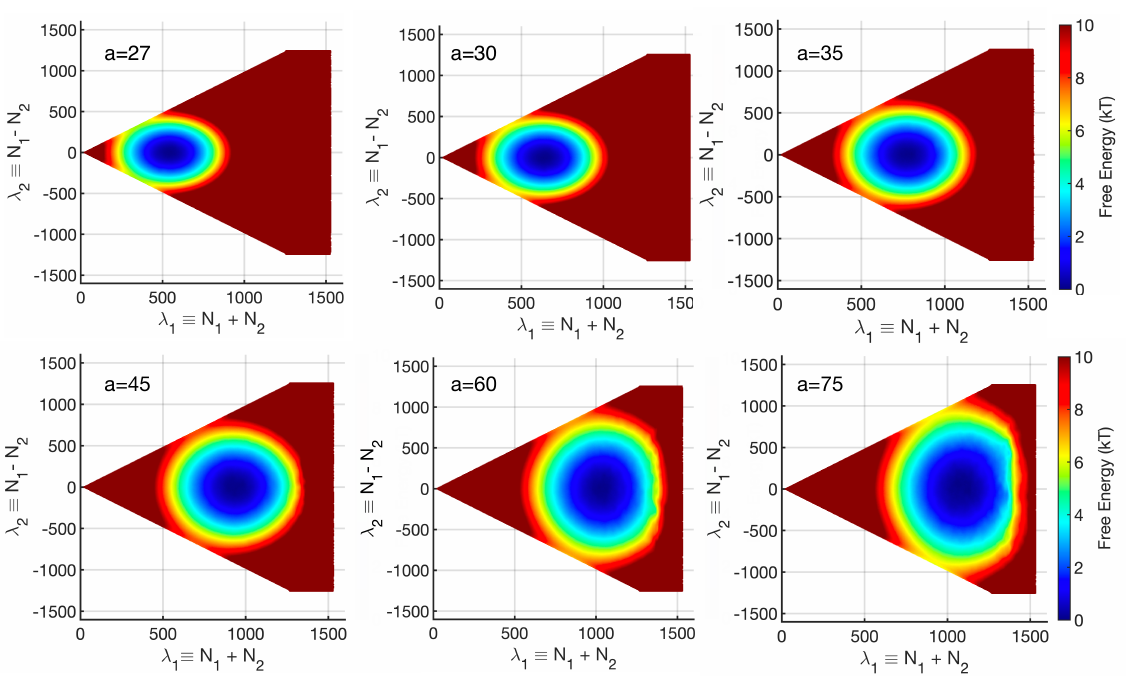}
\end{center}
\caption{Free energy vs $\lambda_1\equiv N_1+N_2$ and $\lambda_2\equiv N_1-N_2$.
Results are shown for a polymer of length $N=1600$ and bending rigidity $\kappa=5$. In addition,
the confining geometry is defined by slit height $h=10$, pit depth $d=20$, and pit spacing $l=60$.
Results are shown for various values of the pit side length, $a$. For visual clarity, the free
energy is truncated at 10$k_{\rm B}T$.}
\label{fig:F2dvarya}
\end{figure*}

In the simulations, we employ semiflexible polymer chains composed of $N$ spherical monomers 
of diameter and fixed bond length both equal to $w$. Note that this is also the width 
of the chain. Since all quantities of distance are given in units of $w$,
$L_1$ and $L_2$ are numerically equal to $N_1$ and $N_2$, the number of monomers
whose centers lie in the nanopits, and $\lambda_1=N_1+N_2$ and $\lambda_2=N_1-N_2$ are the
sum and the difference of the number of monomers in the two pits. Likewise, $N_{\rm s}=N-(N_1+N_2)
=N-\lambda_1$ is the number of monomers that lie outside the pits. Finally, note that the
coefficients in Eqs.~(\ref{eq:Fconf3}) and (\ref{eq:Fsumdiff}) have the following dependencies: 
$A=A(a,d,P,h)$ and $B=B(d,a)$. The dependence on polymer width $w$ is omitted since this
is the unit of length employed in this study.

For fixed $\lambda_1=\lambda_1^{\rm (min)}$, (i.e., $\delta\lambda_1=0$) the free energy varies as
\be
\frac{F_{\rm conf}(\lambda_2)}{k_{\rm B}T} & = & \frac{\kappa_{\rm a}}{2} (\lambda_2)^2,
\label{eq:Fvary2}
\ee
where 
\be
\kappa_{\rm a} = B = \frac{\pi w}{4da^2}.
\label{eq:kappaa}
\ee
However, at $\lambda_2=\lambda_2^{\rm (min)}=0$, 
\be
\frac{F_{\rm conf}(\lambda_1)}{k_{\rm B}T} & = & \frac{\kappa_{\rm s}}{2} (\delta\lambda_1)^2 
+ {\cal O}\left(\delta \lambda_1^3\right).
\label{eq:Fvary1}
\ee
Equations~(\ref{eq:Fvary2}) and (\ref{eq:Fvary1}) will be useful for comparison with 
the simulation data.

\section{Results}
\label{sec:results}

We begin with an examination of the effects of the cavity side length, $a$ on the
free energy functions. Note that this was a key focus in the experimental study of 
Ref.~\onlinecite{klotz2015correlated}.  Figure~\ref{fig:F2dvarya} 
shows a collection of graphs of the free energy, $F(\lambda_1,\lambda_2)$,
{where $\lambda_1$ and $\lambda_2$ are the sum and difference of the
number of spherical monomers in the nanopits, respectively.  Results are shown}
for a system composed of a polymer of length $N=1600$ monomers and persistence length
of $P=5$ confined in a slit of height $h=10$. The two nanopits embedded in one of the
slit surfaces have a depth of $d=20$ and spacing $l=60$. The individual graphs show
results for various values of the nanopit side length, $a$, ranging from $a=30$ to $a=90$.
The qualitative trends are straightforward. In all cases, the free energy minimum is
centered at a point with $\lambda_2=0$, which implies that the most probable configuration
is one in which $N_1=N_2$, i.e., an equal number of monomers located in each pit.
The value of $\lambda_1^{\rm (min)}$ increases gradually with increasing $a$. This
implies that $\lambda_1\equiv N_1+N_2$, the total number of monomers located in the two
pits, increases as the pit width increases. This follows from the fact that larger pits
can more comfortably accommodate larger numbers of monomers since the impact of
the excluded volume interactions, determined by the coefficient $B$, defined in 
Eq.~(\ref{eq:deff2}), decreases as the cavity volume increases.

\begin{figure*}[!htb]
\begin{center}
\includegraphics[width=0.90\textwidth]{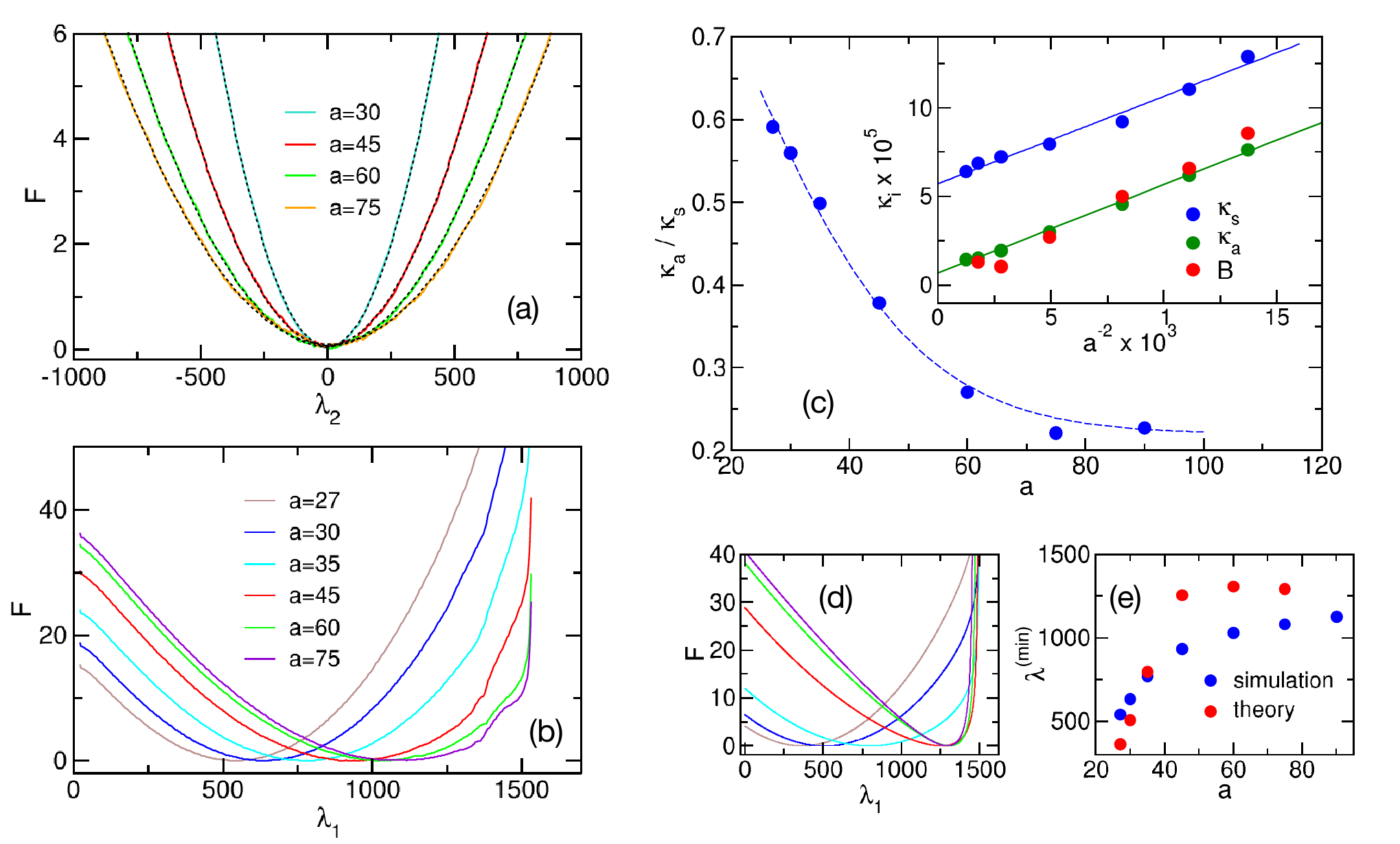}
\end{center}
\caption{ (a) $F$ vs  $\lambda_2$ for fixed $\lambda_1 =\lambda_1^{{\rm (min)}}$.  
The results were obtained from a cross sections of the 2D free energy functions of 
Fig.~\ref{fig:F2dvarya} As in Fig.~\ref{fig:F2dvarya}, $N=1600$, $\kappa=5$, $h=10$, 
$d=20$, and $l=60$.  The dotted lines show quadratic fits to the data. 
(b) $F$ vs $\lambda_1$ for fixed $\lambda_2=0$.  
(c) The inset shows the variation of $\kappa_{\rm s}$ and $\kappa_{\rm a}$ with $a^{-2}$.
Here, $\kappa_{\rm a}$ and $\kappa_{\rm s}$ are the coefficients obtained for fits to
$F(\lambda_2;\lambda_1=\lambda_{1}^{\rm (min)})$ and $F(\lambda_1;\lambda_2=0)$, respectively.
Solid lines show fits using the $F=\gamma_0 + \gamma_1a^{-2}$. The data were obtained from
fits to the data in Fig.~\ref{fig:Flam12}, as described in the text. Overlaid on the data are
estimates for the excluded-volume coefficient $B$ (see Eq.~(\ref{eq:Fbox})) obtained from
calculations of the confinement free energy of a single polymer chain in a rectangular box 
with the same dimensions as the nanopits, as described in the text.  The main part of the 
figure shows the variation of the ratio $\kappa_{\rm a}/\kappa_{\rm s}$ with pit side length 
$a$.  {The dashed line is a guide for the eye and was calculated using the function
$\kappa_{\rm a}/\kappa_{\rm s} = 0.22 + 0.608 e^{-((a-10)/28.39)^{1.5}}$.}
(d) Predicted $F$ vs $\lambda_1$ using Eq.~(\ref{eq:Fconf3}). The colors for the curves
for the values of $a$ are the same as in (b). 
(e) Variation of $\lambda_1^{(\rm min)}$ with $a$ determined from the free energy
functions shown in (b) and the theoretical curves shown in (d).
}
\label{fig:Flam12}
\end{figure*}

Figure~\ref{fig:Flam12}(a) shows the cross section of the free energy through
the free energy minimum along the $\lambda_2$ axis.  As predicted by Eq.~(\ref{eq:Fvary2}),
the free energy exhibits a quadratic 
dependence on $\lambda_2$. Indeed the fits of each curve to the function $F=c_1(\lambda_2)^2$, 
shown as the dotted lines in the figure, are in excellent agreement with the data. 
The effective spring constant can be evaluated using $\kappa_{\rm a}=2c_1$.
Figure~\ref{fig:Flam12}(b) shows the cross section of the free energy through
the free energy minimum along the $\lambda_1$ axis.  The shift in the minimum to higher $\lambda_1$
as the pit width $a$ increases is evident. 
The free energy curves in Fig~\ref{fig:Flam12}(a) were fit to the polynomial,
$F(\lambda_1)=c_2(\lambda_1-\lambda_1^{\rm (min)})^2 + c_3(\lambda_1-\lambda_1^{\rm (min)})^3$,
in the vicinity of the free energy minimum and used to obtain the effective spring constant,
$\kappa_{\rm s} = 2c_2$.

The inset of Figure~\ref{fig:Flam12}(c) shows that the $\kappa_{\rm a}$ values extracted 
from the fits scale roughly linearly with $a^{-2}$.  This scaling is qualitatively 
consistent with the prediction of Eq.~(\ref{eq:kappaa}). In the limit of large pit width 
(i.e., small $a^{-2}$), $\kappa_{\rm a}$ appears to deviate from the linear curve and 
approach a finite value. A fit of these data (excluding the data point for the largest 
$a$ value)  to the function $F=\gamma_0+\gamma_1a^{-2}$ yields 
$\gamma_0=(6.6\pm 0.9)\times 10^{-6}$ and $\gamma_1=0.050\pm 0.001$. By comparison,
Eqs.~(\ref{eq:Fvary2}) and (\ref{eq:kappaa}) predict a linear variation of $F$ with
respect to $a^{-2}$ with zero intercept and slope of $\pi w/(4d)=0.039$. This value
is value is in close agreement with the measured value of $\gamma_1$.

In order to further assess the observed variation of $\kappa_{\rm a}$ with $a^{-2}$,
we carry out a complementary simulation to measure the confinement
free energy of a polymer trapped in a box of dimensions $a\times a\times d$.
We follow the approach taken in a previous study,\cite{polson2019polymer}
which differs from that used in the rest of the study; details of the calculation 
are presented in {Appendix~\ref{app:b}}. The simulations provide an estimate
of the variation of the confinement free energy with respect to polymer length
for a polymer with the same bending rigidity ($\kappa=5$) as used in the two-pit
simulations. For sufficiently long polymers the free energy is expected to vary with 
contour length according to Eq.~(\ref{eq:Fbox}). Fitting the data with 
$F(m)=A+Bm^2$, where $m$ is the polymer length, yields values of the virial 
coefficient, $B$. As noted earlier, the effective spring constant for
the contour-length difference is expected to satisfy $\kappa_{\rm a}=B$.
Overlaid on the data in the inset of Fig.~\ref{fig:Flam12}(c) are the values of $B$
calculated using this approach. The values of $\kappa_{\rm a}$ and $B$ are
in reasonable agreement over the range of $a$ examined. The small discrepancies 
arise mainly from the fact that the nanopit is partially ``open'' in the slit
portion of the geometry in contrast to the fully enclosed box used to estimate $B$.
The results of this calculation show that the theoretical model presented in
Section~\ref{sec:model} accounts for the variation of $F$ with $\lambda_2$.

As noted in Section~\ref{sec:theory}, when the contour length of that portion of
the polymer located outside the nanopits is sufficiently large  (i.e., low $\lambda_1$) 
then the spring free energy of Eq.~(\ref{eq:Fspr}) is negligible. In this case, 
the confinement free energy of Eq.~(\ref{eq:Fsumdiff}) can be approximated by
Eq.~(\ref{eq:Fsea}), implying that $\kappa_{\rm s}=\kappa_{\rm a}$. 
The  variation of the measured value of $\kappa_{\rm s}$ with $a^{-2}$ shown
in the inset of Fig.~\ref{fig:Flam12}(c) clearly shows that this is not correct.
Instead, $\kappa_{\rm s}$ is consistently greater than $\kappa_{\rm a}$ for all $a$. 
Restricting the fit to smaller values of $\lambda_1$ well below $\lambda_1^{\rm (min)}$
where the spring free energy $F_{\rm spr}$ of Eq.~(\ref{eq:Fspr}) is expected to
be negligible does not significantly change the results.  Like $\kappa_{\rm a}$, 
the coefficient $\kappa_{\rm s}$ does appear to vary roughly linearly with $a^{-2}$, 
but the two coefficients are offset by roughly 
$\Delta\kappa\equiv \kappa_{\rm s}-\kappa_{\rm a}\approx 5\times 10^{-5}$ 
for the entire range of $a$ examined here.  The main part of Fig.~\ref{fig:Flam12}(c) 
shows the variation of the ratio $\kappa_{\rm a}/\kappa_{\rm s}$ with respect to pit 
side length $a$. The ratio decreases rapidly at low $a$ and appears to asymptotically 
approach $\approx 0.2$ at larger $a$.
Figure~\ref{fig:Flam12}(d) shows the predicted variation of $F$ with $\lambda_1$
(for $\lambda_2=0$) using Eq.~(\ref{eq:Fconf3}). The qualitative trends are
comparable to those in Fig.~\ref{fig:Flam12}(b), most notably the shift in the
location of the minimum, $\lambda_1^{(\rm min)}$, to higher values as the
cavity width increases. However, these values tend to underestimate the measured
values at low $a$ and overestimate them at higher $a$, as is clearly shown in 
Fig.~\ref{fig:Flam12}(e). 

It is useful to compare our simulation results with those of 
Ref.~\onlinecite{klotz2015correlated}. Note that the experiments and simulations
in that work did not measure the free energy directly as in the present study, but
rather analyzed the dynamics of the fluctuations in $\lambda_1$ and $\lambda_2$.
However, the correlation times for the fluctuations are closely related to the two effective 
spring constants. The spring constant $\kappa_{\rm a}$ governs the fluctuations in the 
back-and-forth trading of polymer contour between the two nanopits. As noted in
Ref.~\onlinecite{klotz2015correlated} the correlation time in the fluctuations is expected
to satisfy $\tau_{\rm a} = \xi/\kappa_{\rm a}$, where $\xi$ is the hydrodynamic drag on
the chain. This implies that $\tau_{\rm a} \propto a^2$. Klotz {\it et al.} measured
$\tau_{\rm a}$ for $\lambda$ DNA in the same complex confinement geometry modeled in
the present work, and the data are roughly consistent with the predicted quadratic dependence
of $\tau_{\rm a}$ with $a$ (see Fig.~5(a) of Ref.~\onlinecite{klotz2015correlated}).

The spring constant $\kappa_{\rm s}$ governs the fluctuations in the total contour trapped in 
the two pits, whose correlation time is expected to vary as $\tau_{\rm s}=\xi/\kappa_{\rm s}$.
Consequently, the ratio of correlation times should satisfy
$\tau_{\rm s}/\tau_{\rm a}=\kappa_{\rm a}/\kappa_{\rm s}$. 
The ratio of $\kappa_{\rm a}/\kappa_{\rm s}$
is roughly consistent with the measured variation of $\tau_{\rm s}/\tau_{\rm a}$ for $\lambda$ DNA
(see Fig.~5(c) in Ref.~\onlinecite{klotz2015correlated}). At $a=30$ the ratio curvature-constant 
ratio is about $\kappa_{\rm a}/\kappa_{\rm s}\approx 0.56$ and appears to level off at 
around $a\approx 60$. For DNA with an effective width of 10~nm, these correspond to
$a=300$~nm and $a=600$~nm, respectively. In Ref.~\onlinecite{klotz2015correlated} 
the correlation-time ratio $\tau_{\rm s}/\tau_{\rm a}$ is about 0.62 at $a=300$~nm, and 
the curve appears to level off at about $a=500$~nm. Thus, there is decent quantitative 
agreement between our simulations and experiment. On the other hand, $\kappa_{\rm a}/\kappa_{\rm s}$ 
appears to approach approximately 0.22 at large $a$, while the limiting value in the experiments 
is just below 0.1. The quantitative discrepancies could arise from the implicit neglect of
the ${\cal O}(\lambda^3)$ correction terms in the approximation
$\kappa_{\rm a}/\kappa_{\rm s}=\tau_{\rm s}/\tau_{\rm a}$. 
In addition, note the contour length of the model polymer chain of $N=1600$ corresponds to a
contour length of $L_{\rm c}=1600w$. For an effective width of $w=10$~nm, this corresponds to 
$L_{\rm c}$=16~$\mu$m, slightly shorter than the value of $L_{\rm c}$=16.5~$\mu$m of $\lambda$ DNA
and even shorter still than the contour length of $\lambda$ DNA stained with YOYO-1 dye at
a 10:1 base-pair/dye ratio, which is expected to increase $L_{\rm c}$ by roughly 
15\%.\cite{kundukad2014effect} The small but significant difference between the experimental 
and simulation values of $L_{\rm c}/w$ may also account for the discrepancies in the data.

The simulations carried out in Ref.~\onlinecite{klotz2015correlated} used a semiflexible
chain of spherical beads interacting with the Weeks-Chandler-Anderson (WCA) particles
connected by a finitely extensible nonlinear elastic spring potential with model parameters
set as in Ref.~\onlinecite{grest1986molecular}. In addition a three-particle harmonic
potential was used to impose bending rigidity with a persistence length of $P\approx 5w$. 
Although the details differ this model is very similar in effect as the semiflexible
hard-sphere chain in the present study. On the other hand, to make the simulations computationally
feasible, the contour length ($L_{\rm c}/w\approx N$=100--300~monomers) was significantly 
shorter than the value here ($L_{\rm c}/w=N$=1600~monomers), and shorter still than the 
experimental value.  Likewise, the values of the various confinement geometry parameters 
(slit height and cavity depth, side length, and separation distance) were also significantly 
smaller (measured in units of $w$) than the experimental values, as noted by the authors of 
the study. In spite of these limitations, variation of the values of $\tau_{\rm s}/\tau_{\rm a}$ 
with respect to $a$ measured in the simulations were in decent quantitative agreement
with the experimental values and the theoretical predictions of Eq.~(\ref{eq:Fconf3}),
as well as the results of our simulations.  We speculate that the choice of the
confinement dimensions to compensate for the artificially short polymer chains
used in the simulations of Ref.~\onlinecite{klotz2015correlated} in effect produced a 
free-energy landscape comparable to that calculated for our model as well as the one 
for the experimental system. What has emerged from our simulations, however, is
that this cannot be taken as corroboration of the theoretical model for the free
energy whose predictions yield such significant quantitative discrepancies with 
our results.

\begin{figure*}[!htb]
\begin{center}
\vspace*{0.2in}
\includegraphics[width=0.90\textwidth]{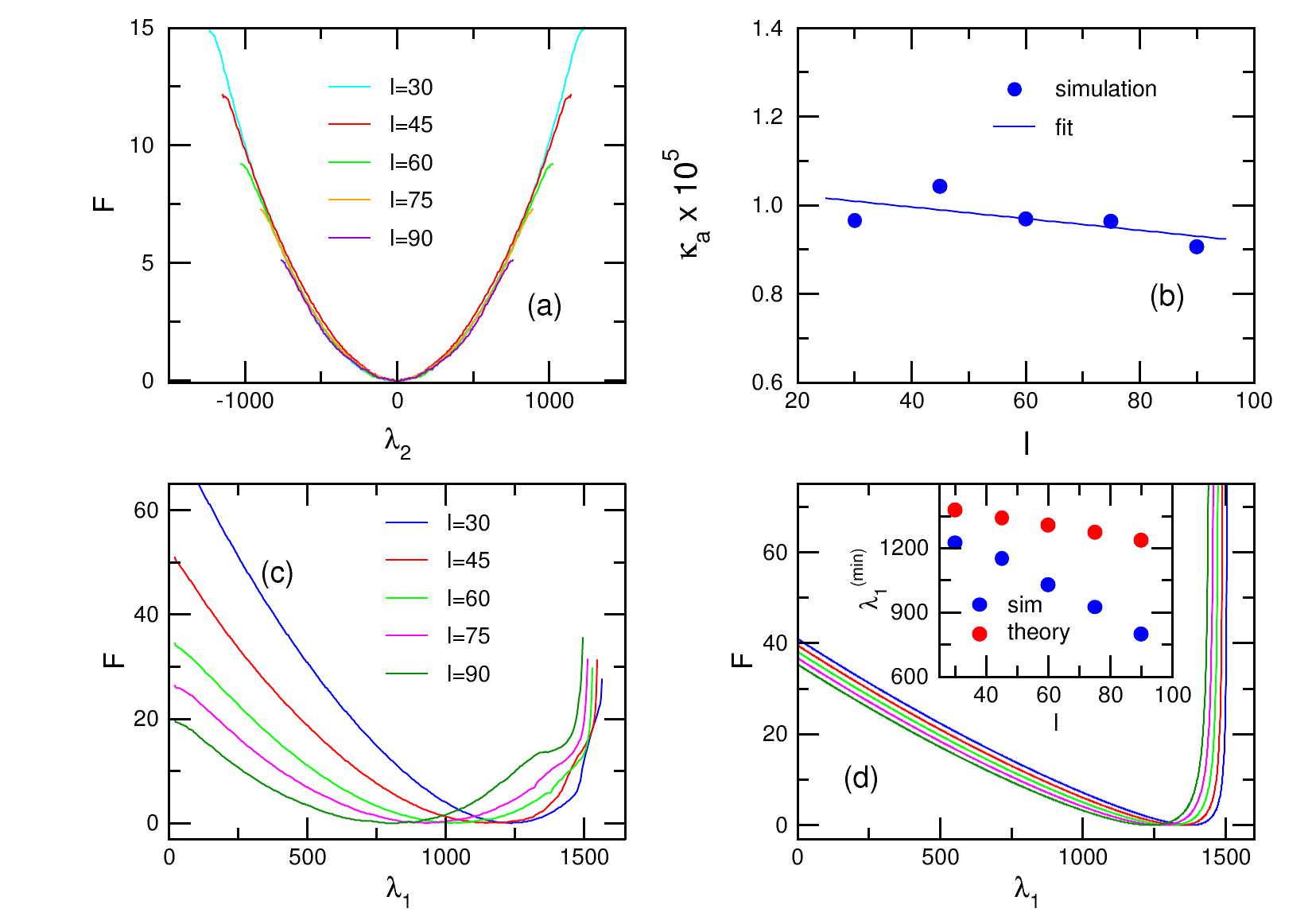}
\end{center}
\caption{Effects of varying the cavity separation, $l$. Results are shown for 
fixed $a=60$, $h=10$, $d=20$, $\kappa=5$, and for several values of $l$.
(a) Free energy vs $\lambda_2$ for fixed $\lambda_1=\lambda_1^{\rm (min)}$.
(b) Variation of $\kappa_{\rm a}$ with $l$. The
values of $\kappa_{\rm a}$ were obtained from fits to the curves in (a).
(c) Free energy vs $\lambda_1$ for fixed $\lambda_2=0$.
(d) Prediction of $F$  vs $\lambda_1$ using Eq.~(\ref{eq:Fconf3}). The
inset shows the variation of $\lambda_1^{\rm (min)}$ with $l$ for
the free energies obtained from simulations and the theory.  }
\label{fig:Flam12_vary_l}
\end{figure*}

The other geometric parameter that was varied in the experiments of 
Ref.~\onlinecite{klotz2015correlated} was the spacing $l$ between the two
nanopits. Klotz {\it et al.} found that the asymmetric mode correlation
time, $\tau_{\rm a}$ increased linearly with $l$ with a rate that increased
with the nanopit width, $a$. This is consistent with a prediction that
assumes $\tau_{\rm a}=\xi/\kappa_{\rm a}$, where $\kappa_{\rm a}=w/a^2d$ is
independent of $l$, and that the friction coefficient satisfies $\xi=\alpha\eta l + \epsilon$,
where $\alpha$ is a numerical proportionality factor, $\eta$ is the buffer viscosity
and $\epsilon$ is an offset factor. The latter assumption invokes the blob model
for the linking strand, with the result that the stretched linking strand length
is proportional to $l$. While our MC simulations cannot be used to evaluate the validity
of the assumed behavior of $\xi$, they can be used to test the assumed independence
of $\kappa_{\rm a}$ to $l$.  Figure~\ref{fig:Flam12_vary_l}(a) shows the $F$ vs $\lambda_2$
free energy cross sections for various $l$ with the other geometric parameters fixed
to $a=60$, $d=20$ and $h=10$. As in Fig.~\ref{fig:Flam12}(a), the functions vary
quadratically with $\lambda_2$. Fitting the data yields the effective spring constant
$\kappa_{\rm a}$ for each curve. Figure~\ref{fig:Flam12_vary_l}(b) shows the variation
of $\kappa_{\rm a}$ with $l$ over the range $l=30-90$. A linear fit of the data
yields a slope of $(1.3\pm 0.8)\times 10^{-8}$. Thus, our results do not show
any statistically significant variation in $\kappa_{\rm a}$ with cavity spacing,
thus validating the assumption employed in the analysis of the experimental data.

Figure~\ref{fig:Flam12_vary_l}(c) shows the effect of varying $l$ on the
$F$ vs $\lambda_1$ free energy cross sections. The main trend is the shift
of $\lambda_1^{\rm (min)}$ to lower $\lambda_1$ as the cavity spacing
increases. This follows trivially from the fact that increasing $l$ 
increases the linking strand length and so decreases the total polymer contour 
that available to occupy the two cavities. Figure~\ref{fig:Flam12_vary_l}(d)
shows the predicted curves using Eq.~(\ref{eq:Fconf3}). While the reduction
in $\lambda_1^{\rm (min)}$ with increasing $l$ is observed, the trend is much 
weaker in the prediction. This is especially clear in the inset of the figure
which compares the simulation and theoretical values of $\lambda_1^{\rm (min)}$. 
Thus, as was the case for the variation of the cavity width, Eq.~(\ref{eq:Fconf3})
yields qualitatively correct but quantitatively poor predictions for $F$ vs $\lambda_1$,
while yielding quantitatively accurate results for $F$ vs $\lambda_2$.

Although the experimental study of Ref.~\onlinecite{klotz2015correlated} characterized 
the trends of varying only the cavity width and spacing, it is of interest here to examine 
the effects of varying other system parameters. We choose the cavity depth parameter $D$, 
slit spacing $h$, as well as the polymer persistence length, $P$. Our results serve as 
predictions for any future experiments that may examine such effects. In addition, they provide 
further tests for the predictions of the theoretical model of Section~\ref{sec:theory}.

\begin{figure*}[!htb]
\begin{center}
\vspace*{0.2in}
\includegraphics[width=0.90\textwidth]{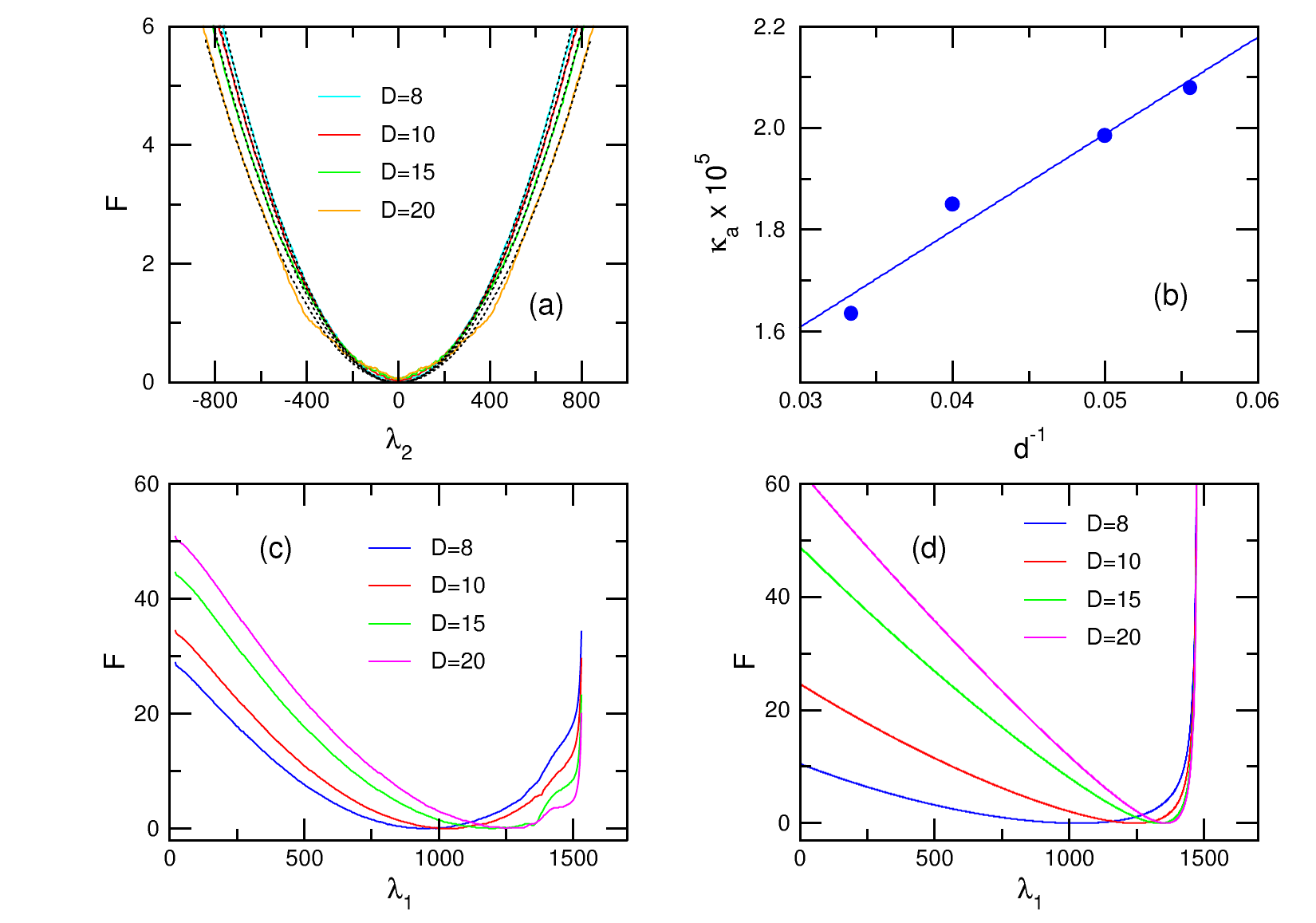}
\end{center}
\caption{Effects of varying the cavity depth, $D$. Results are shown for 
fixed $a=60$, $h=10$, $l=60$, $\kappa=5$, and for several values of $D$.
(a) Free energy vs $\lambda_2$ for fixed $\lambda_1=\lambda_1^{\rm (min)}$.
(b) Variation of $\kappa_{\rm a}$ with $d^{-1}\equiv (h+D)^{-1}$. The
values of $\kappa_{\rm a}$ were obtained from fits to the curves in (a).
(c) Free energy vs $\lambda_1$ for fixed $\lambda_2=0$.
(d) Prediction of $F$  vs $\lambda_1$ using Eq.~(\ref{eq:Fconf3}).  }
\label{fig:Flam12_vary_D}
\end{figure*}

We consider first the effect of changing $D$, the pit depth as measured from the embedding 
surface. Figure~\ref{fig:Flam12_vary_D}(a) shows  free energy cross sections $F(\lambda_2)$
for several values of $D$ with fixed $h=10$, $a=60$, and $l=60$.
The dotted lines are parabolic fits to $F(\lambda_2)$, from which the coefficient
$\kappa_{\rm a}$ has been extracted. The parabolas in Fig.~\ref{fig:Flam12_vary_D}(a)
widen slightly with increasing $D$. Correspondingly, $\kappa_{\rm a}$ increases
monotonically with $d^{-1}$, where $d=h+D$, as shown in Fig.~\ref{fig:Flam12_vary_D}(b). 
A linear fit of $\kappa_{\rm a}$ vs $d^{-1}$ yields a slope of $(1.9\pm 0.3)\times 10^{-4}$.
Equation~(\ref{eq:kappaa}) predicts a linear variation with a slope 
$\pi w/(4a^2)= 2.2\times 10^{-4}$, which is comparable to the measured value.

Figure~\ref{fig:Flam12_vary_D}(c) shows the effect of varying $D$ (through variation of
$D$ at fixed $h$) on the cross section $F(\lambda_1)$. The principal
effect is that the minimum value $\lambda_1^{\rm (min)}$ increases with $D$. This effect
is due to the fact that increasing $d$ increases that cavity volume, thus making
it easier to accommodate a greater portion of the polymer contour. This is exactly
the same explanation for the qualitative trend for varying $a$ in Fig.~\ref{fig:Flam12}(b).
As in that case, however, shift of the curves to lower $\lambda_1$ as the cavity volume
increases are only qualitatively consistent with the theoretical predictions shown
in Fig.~\ref{fig:Flam12_vary_D}(d). 

\begin{figure*}[!htb]
\begin{center}
\vspace*{0.2in}
\includegraphics[width=0.90\textwidth]{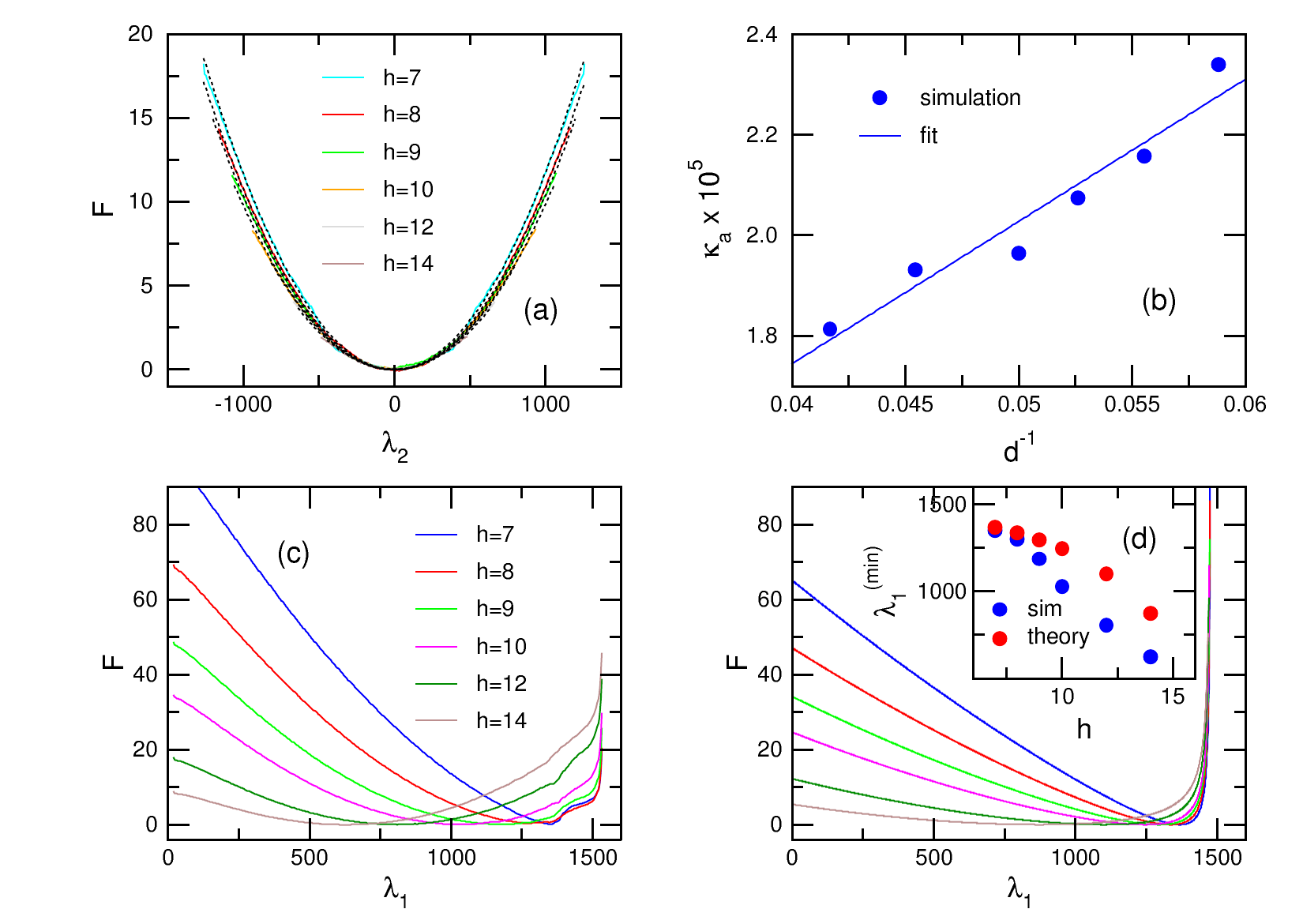}
\end{center}
\caption{Effects of varying the slit height, $h$.  Results are shown for $a=60$, 
$D=10$, $l=60$, and  $\kappa=5$. (a) $F$ vs $\lambda_2$ for $\lambda_1=\lambda_1^{\rm (min)}$. 
The dotted lines are quadratic function fits to the calculated curves. 
(b) Variation of $\kappa_{\rm a}$, extracted from the fits in (a), with respect to $d^{-1}$,
where $d=h+D$. The dashed line is a linear fit to the data.  
(c) $F$ vs $\lambda_1$ for $\lambda_2=0$. 
(d) Prediction of $F$  vs $\lambda_1$ using Eq.~(\ref{eq:Fconf3}).  The
inset shows the variation of $\lambda_1^{\rm (min)}$ with $h$ for
the free energies obtained from simulations and the theory.  }
\label{fig:Flam12_vary_h}
\end{figure*}

Another way of varying $d$, and thus the pit volume, is adjusting the slit height
$h$ while keeping the cavity depth $D$ fixed.  Figure~\ref{fig:Flam12_vary_h}(a) 
shows the effect of such a variation on the cross section $F(\lambda_2)$. Increasing
$h$ widens the parabolas. A fit to parabolic functions provides estimates of the
coefficient $\kappa_{\rm a}$, which is plotted in Fig.~\ref{fig:Flam12_vary_h}(b)
as a function of $d^{-1}$. A fit of the data yields a slope of $(2.8\pm 0.4)\times 10^{-4}$, 
which is in good agreement with the prediction from Eq.~(\ref{eq:kappaa}) of 
$\pi w/(4a^2)= 2.2\times 10^{-4}$.
Figure~\ref{fig:Flam12_vary_h}(c) shows the effect of varying the slit height on the
cross section $F(\lambda_1)$. Increasing $h$ reduces $\lambda_1^{\rm (min)}$,
the most probable value of the total polymer contour in the pits.
The theoretical predictions for $F$ vs $\lambda_1$ shown in Fig.\ref{fig:Flam12_vary_h}(d)
again show the correct qualitative trend, in this case decreasing $\lambda_1^{\rm (min)}$
with increasing $h$, though with significant quantitative discrepancies with the
curves obtained from the simulations.

\begin{figure*}[!htp]
\begin{center}
\vspace*{0.2in}
\includegraphics[width=0.90\textwidth]{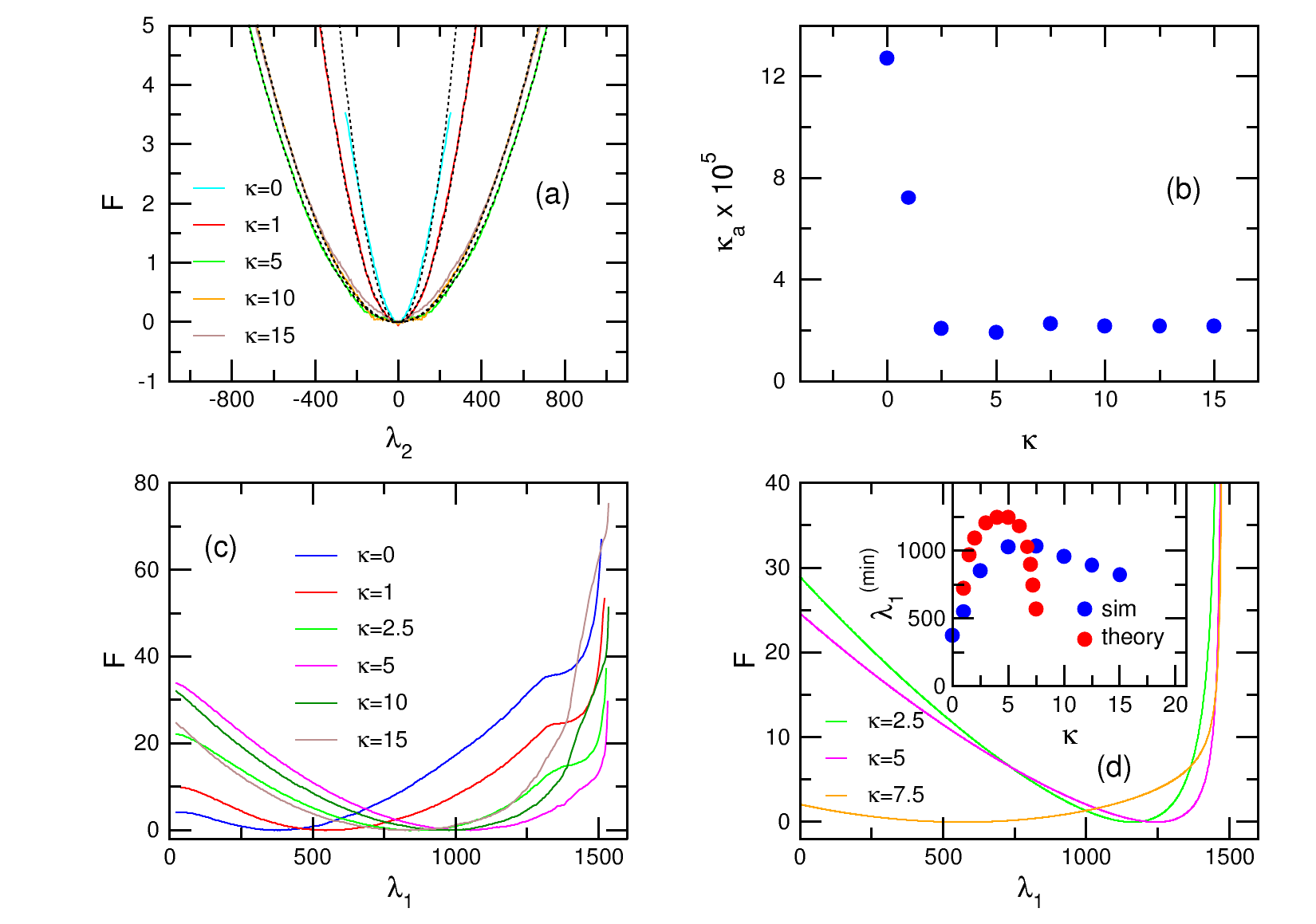}
\end{center}
\caption{Effects of the variation of the polymer bending rigidity, $\kappa$.
{Note that increasing $\kappa$ corresponds physically to increasing
the persistence length.}
Results are shown for fixed $a=60$, $d=20$, $h=10$ and $l=60$.
(a) $F$ vs $\lambda_2$ for $\lambda_1=\lambda_1^{(\rm min)}$. The dotted
curves are quadratic fits to the data. (b) Effective spring constant, $\kappa_{\rm a}$
vs bending rigidity, where $\kappa_{\rm a}$ have been determined from the fits
to the data in (a). (c) Variation of $F$ with $\lambda_1$. 
(d) Prediction of $F(\lambda_1)$ using Eq.~(\ref{eq:Fconf3}). The inset shows
the variation of $\lambda_1^{\rm (min)}$ with $\kappa$ for both the simulation
results and the theoretical predictions.}
\label{fig:Flam12_vary_P}
\end{figure*}

Finally, we examine the effects of varying the polymer persistence length on 
the free energy functions. Figure~\ref{fig:Flam12_vary_P} shows results for
varying the polymer bending rigidity, $\kappa$, for a system with $a=60$, 
$d=20$ and $h=10$. As in previous calculations, the $F$ varies quadratically 
with $\lambda_2$, as shown in Fig.~\ref{fig:Flam12_vary_P}(a). Values of
the effective spring constant $\kappa_{\rm a}$ were obtained from fitting
the data and are shown in Fig.~\ref{fig:Flam12_vary_P}(b). There is a rapid
initial reduction in $\kappa_{\rm a}$ until about $\kappa\approx 2.5$, after
which $\kappa_{\rm a}$ remains relatively constant.  The invariance of $\kappa_{\rm a}$
with respect to $\kappa$ is consistent with the prediction of Eq.~(\ref{eq:kappaa}), 
which has no dependence on $P$. The deviation from this result at very low $\kappa$
may be due to the neglect of higher order terms in the virial expansion of $F_{\rm box}$
(see Eq.~(\ref{eq:Fbox})), which are expected to become more significant with
increasing the polymer flexibility.

As is evident in Fig.~\ref{fig:Flam12_vary_P}(c), varying the bending rigidity has
a pronounced effect on the $F$ vs $\lambda_1$ curves. Interestingly, the position
of the minimum does not vary monotonically with $\kappa$. As is clearer in the
inset of Fig.~\ref{fig:Flam12_vary_P}(d), $\lambda_1^{\rm (min)}$ initially increases 
with $\kappa$, reaches a maximum at about $\kappa=5$ and then decreases. 
Figure~\ref{fig:Flam12_vary_P}(d) shows theoretical predictions calculated using
Eq.~(\ref{eq:Fconf3}). Only curves for $\kappa$=2.5, 5 and 7.5 are shown. For
$\kappa \gtrsim 8$ the theory fails completely as the coefficient $A$ defined in
Eq.~(\ref{eq:Adef}) becomes negative because $A_{\rm slit} < A_{\rm box}$,
where these two quantities are defined in Eq.~(\ref{eq:deff2CS}) and Eq.~(\ref{eq:deff1box}),
respectively. A likely cause of this problem is the expression for $A_{\rm box}$ 
in Eq.~(\ref{eq:deff1box}) is strictly only valid for $P\ll d$ and $P\ll a$, conditions
that break down as the bending rigidity increases.

\section{Conclusions}
\label{sec:conclusions}

In this study we have used MC simulations to measure a 2D free-energy landscape for
a polymer confined to a complex geometry consisting of a slit with two cavities
embedded in one of the surfaces. The simulation results were used to critically
evaluate the theoretical model used to interpret the results of an experiment
that examined the physical behavior of a $\lambda$ DNA molecule confined to this
geometry.\cite{klotz2015correlated} The values of parameters that define the lengthscales 
in the simulations for both the confining geometry and the polymer were chosen to be
close to the lengthscales in the experiments. The free energy was calculated as a
function of the sum, $\lambda_1$, and difference, $\lambda_2$, of the polymer contour
length in the two cavities, and for convenience the analysis focused on the cross 
sections of $F$ along one variable through the minimum of the other. The theoretical
model is constructed using established results for the confinement free energy of
polymers in cavities and slits. 

We observe a marked difference in the accuracy 
of the theoretical predictions for the $\lambda_1$ and $\lambda_2$ cross sections.
We find the predictions for $F(\lambda_2)$ are quantitatively accurate upon
variation of every confining geometry lengthscale, as well as the polymer persistence
length. By contrast, we find only qualitative agreement for $F(\lambda_1)$ and
appreciable quantitative discrepancies.
The origin of the quantitative discrepancy for the case of $\lambda_1$ in
part lies in the simplistic, if convenient, modeling of the surface-embedded 
cavities as enclosed boxes. In most of the simulations (as well as in the experiments)
half of the cavity height (i.e. the distance from the cavity bottom surface to
the other slit surface) is open to the slit such that the polymer is not completely
confined in the lateral dimension. Further, the simple analytical expression for
the cavity confinement free energy likely exhibits finite-size effects for cavities
as small as employed here. For example, it was noted that $F_{\rm box}$ fails 
for longer persistence lengths because of the requirement that $P$ be much smaller
than the cavity dimensions. The failure of the theoretical model to provide
quantitatively accurate predictions for $\lambda_1$ is less surprising than its
success in doing so for $\lambda_2$. Apparently the effects of some of the model 
inaccuracies cancel out for this case. 

Our study provides a useful lesson on the limitations of using simplistic (if convenient)
analytical expressions for polymer free-energy landscapes to interpret experimental
data and points to the value of carrying out accurate (if time consuming) calculations
of the free energy instead. The apparent agreement of the experimental (and even Langevin
dynamics simulation) results of Ref.~\onlinecite{klotz2015correlated} with the theoretical
predictions suggested a theory that was much better than was revealed in our study.
In future work, we will carry out similar calculations of free-energy landscapes
for comparable complex geometries that have been employed in other experiments using
DNA.\cite{nykypanchuk2002brownian, mikkelsen2011pressure, vestergaard2016transition, 
smith2017photothermal, del2009pressure, klotz2012diffusion, klotz2015correlated, 
klotz2016waves, kim2017giant} Such calculations should contribute to the future 
development of confinement-based nanotechnologies for the manipulation and analysis 
of DNA and other polymers.

\begin{acknowledgments}
This work was supported by the Natural Sciences and Engineering Research Council of Canada 
(NSERC).  We are grateful for the advanced computing resources provided by the Digital Research 
Alliance of Canada as well as ACENET, their regional partner in Atlantic Canada. Finally,
we would like thank Alex Klotz for helpful discussions.
\end{acknowledgments}

\appendix

{
\section{Implementation of the WHAM2D method: further details}
\label{app:a}

As noted in Section~\ref{sec:methods}, the WHAM2D method is used to calculate the
two-dimensional free-energy function $F(\lambda_1,\lambda_2)$ for the confined-polymer
model of this study. In this appendix we provide further detail regarding the implementation
of the method.

The free energy is measured as a function of the sum ($\lambda_1$) and the difference ($\lambda_2$)
of the polymer contour contained within each of the two nanopits of the confinement geometry.
The function can be written $F(\lambda_1,\lambda_2)=-k_{\rm B}T\ln{\cal P}(\lambda_1,\lambda_2)$,
where ${\cal P}(\lambda_1,\lambda_2)$ is the probability distribution associated with the two
variables. While a single simulation can in principle be used to estimate ${\cal P}(\lambda_1,\lambda_2)$
(and therefore also $F(\lambda_1,\lambda_2)$), regions of high free energy have a low probability
and are therefore subject to infrequent sampling and thus poor statistical accuracy.

Multiple-histogram umbrella sampling methods such as WHAM2D offer a straightforward means to 
overcome this problem. The basic idea is to carry out a collection of simulations, each of
which employs a unique biasing potential designed to ensure that the system samples a particular
region of the free-energy landscape. WHAM2D uses two-dimensional harmonic potential energy
functions of the following form:
\begin{eqnarray}
U_{\rm b}={\textstyle{\frac{1}{2}}}k_1 (\lambda_1-\lambda_{1}^{\rm (c)})^2
+ {\textstyle{\frac{1}{2}}}k_2 (\lambda_2-\lambda_{2}^{\rm (c)})^2,
\end{eqnarray}
where the biasing energy minimum is located at $\lambda_1=\lambda_{1}^{\rm (c)}$ and 
$\lambda_2=\lambda_{2}^{\rm (c)}$. (Here we employ the notation relevant to the model 
system of the present study.) The system tends to sample regions near the energy minimum
over a range that depends on the biasing potential constants $k_1$ and $k_2$. Sampling 
$\lambda_1$ and $\lambda_2$ yields a two-dimensional histogram for the biased system.
When using the WHAM2D method, a collection of histograms, each calculated using a biasing
potential with a unique choice of $\lambda_{1}^{\rm (c)}$ and $\lambda_{2}^{\rm (c)}$,
is calculated. The WHAM2D algorithm uses these histograms to reconstruct the underlying
unbiased probability distribution, ${\cal P}(\lambda_1,\lambda_2)$. (In practice, the
algorithm uses as input the histories of $\lambda_1$ and $\lambda_2$ that are produced
by the MC program, from which such histograms are calculated.) The details of the 
histogram reconstruction algorithm can be found elsewhere.\cite{kumar1995multidimensional}

A key consideration in the implementation of the method are the choices of the 
parameters $k_1$ and $k_2$, as well as the ``location'' of the biasing potential,
defined by $\lambda_1^{\rm (c)}$ and $\lambda_2^{\rm (c)}$. 
In order for the method to yield statistically sound results, it is essential that 
there be sufficient overlap of neighbouring histograms, i.e., those for nearest-neighbour
locations of biasing potentials. This imposes a relationship between the spacing
of the biasing potential locations and the parameters $k_1$ and $k_2$. 
The biasing potential locations are chosen to be laid out on a rectangular grid
in the $\lambda_1-\lambda_2$ in a region over which the free energy is to be evaluated. 

Figure~\ref{fig:wham2d}(a) illustrates this approach for the confined-polymer model system.
The triangular region is the domain over which the free energy is evaluated. (As noted
in Sec.~\ref{sec:methods}, the definitions of $\lambda_1$ and $\lambda_2$ require
that $\lambda_2\in[-\lambda_1,\lambda_1]$, yielding such a triangular domain.)
The evenly spaced circles in the figure represent equipotential lines for each biasing
potential employed in the calculation. The centers of the circles (i.e. the 
centers of the potentials) are given by $\lambda_1^{\rm (c)}$ and $\lambda_2^{\rm (c)}$. 
The rectangular grid of potential centers are chosen only to lie in the domain of interest 
in the $\lambda_1-\lambda_2$ plane. The spacings $\Delta \lambda_1$ and $\Delta \lambda_2$
along the $\lambda_1$ and $\lambda_2$ axes, respectively, are illustrated in the figure.

\begin{figure}[!htb]
\begin{center}
\vspace*{0.2in}
\includegraphics[width=0.48\textwidth]{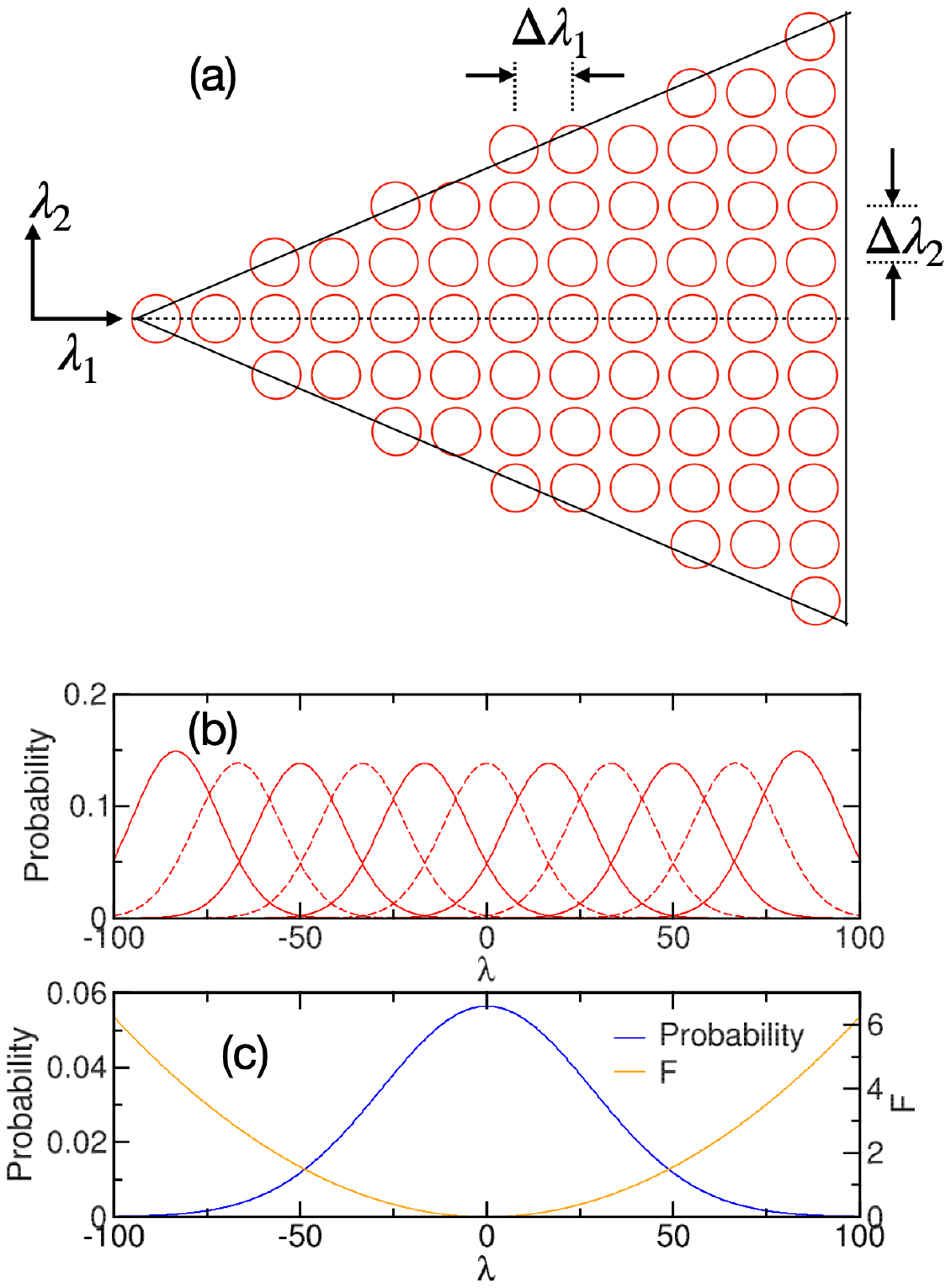}
\end{center}
\caption{{
(a) Schematic illustration of the implementation of the WHAM2D method for the model 
system. The triangle defines the region in the $\lambda_1-\lambda_2$ plane over which 
the free energy is to be evaluated.  Each red circle is associated with a two-dimensional
harmonic biasing potential centered at the circle center, and represents a equipotential 
line for an energy of the order of 1~$k_{\rm B}T$.  
(b) 1D illustration of the WHAM technique. The figure shows a collection of overlapping 
histograms, each of which has been calculated in a simulation using a harmonic 
biasing potential centered at a unique position along the $\lambda$ axis. Curves alternate 
between solid and dashed for visual clarity.
(c) Probability histogram reconstructed from the histograms in (b) using the WHAM 
algorithm. The orange curve is the corresponding free energy profile, 
$F(\lambda)/k_{\rm B}T = -\ln {\cal P}(\lambda)$.}  }
\label{fig:wham2d}
\end{figure}

Figure~\ref{fig:wham2d}(b) and (c) provides a visual description of the method
using a one-dimensional example, for visual clarity. Figure~\ref{fig:wham2d}(b)
shows a collection of histograms calculated for a set of harmonic biasing potentials
evenly spaced along the $\lambda$ axis over a chosen range of $\lambda$. The
potential spacing and the curvature constant of the biasing potential is chosen
to ensure that the distributions overlap appreciably, an essential requirement for
the WHAM and WHAM2D methods. The resulting distributions are used as input into
the WHAM algorithm, which generates the probability histogram for the unbiased
system shown as the blue curve in Fig.~\ref{fig:wham2d}(c). The corresponding 
free-energy function, shown as the orange curve, is determined simply by
$F(\lambda)/k_{\rm B}T = -\ln {\cal P}(\lambda)$. Although Figs.~\ref{fig:wham2d}(b) and 
(c) are 1D illustrations, the case for 2D is essentially the same. In that case,
though, there is freedom to choose biasing potential curvature constants ($k_1$
and $k_2$) and lattice spacings for the biasing potentials ($\Delta\lambda_1$
and $\Delta\lambda_2$) that differ in the two dimensions.

}

\section{Calculation of a semiflexible polymer confined to a rectangular cavity}
\label{app:b} 

In Section~\ref{sec:results}, we examined the variation of the effective spring 
constant, $\kappa_{\rm a}$, with the pit side length $a$. As noted in
Section~\ref{sec:theory} the spring constant is expected to satisfy $\kappa = B$,
where $B$ is the virial coefficient appearing in Eq.~(\ref{eq:Fbox}) that accounts
for the excluded-volume interactions in a semiflexible polymer confined to a rectangular
box of dimension $a\times a\times d$. The coefficient $B$ was calculated by measuring
the variation of the free energy of such a confined polymer with respect to its contour length.
The results for $B$, shown in Fig.~\ref{fig:Flam12}(c) agree reasonably well with those for
$\kappa_{\rm a}$. In this appendix we present the method used in this calculation.

We follow the approach taken in Ref.~\onlinecite{polson2019polymer}. In essence, we
measure the variation of the conformational free energy of a polymer that translocates
through a narrow hole in a hard wall from a ``semi-free'' space on one side ({\it cis})
of the wall to a hard-walled box of dimensions $a\times a\times d$ on the other side ({\it trans}). 
In the case where $m$ monomers of a $N$-monomer chain have translocated into the box, the 
free energy is $F_1(m) = F_{\rm trans}^{(cw)}(m) - F_{\rm cis}^{(w)}(N-m)$. The first
term denotes the free energy of a chain of length $m$ tethered to a wall (``w'') and 
confined to a cavity (``c'') on the {\it trans} side of the pore, i.e., into the cavity.
A polymer that translocates through a pore in a barrier with no confinement on either
side likewise has a free energy $F_0(m) = F_{\rm trans}^{(w)}(m) - F_{\rm cis}^{(w)}(N-m)$.
We define the confinement free energy as $F_{\rm conf}(m) = F_1(m)-F_0(m) = F_{\rm trans}^{(cw)}(m)-
F_{\rm trans}^{(w)}(m)$. Thus, calculation of two translocation free energy functions, one with 
and one without cavity confinement on the {\it trans} side yields the confinement free 
energy.  Note that this represents the change in the free energy required to confine a chain
tethered at one end to a flat wall to cavity. However, as we have noted in a previous
study\cite{polson2019free} the free energy cost of localizing an end monomer to interior
surface of a confining cavity is small. It is not expected to affect the calculation of $B$.

Translocation free energy functions for a chain of length $N=$500 and bending rigidity 
$\kappa=5$ were calculated with and without a rectangular confining cavity on one side,
and the difference between the two was used to calculate $F_{\rm conf}(m)$. As in 
Ref.~\onlinecite{polson2019polymer} each free energy calculation employed the 
Self-Consistent Histogram (SCH) method.\cite{frenkel2002understanding} This is essentially
the equivalent of the 1D WHAM method, except that it uses square-well biasing potentials 
(``windows'') instead of quadratic functions. A set of 199 simulations were carried out,
each a biasing potential window width of $\Delta m=5$, centered on $m$ values such that
each window overlapped by half with adjacent windows. The simulations were each run
for $5\times 10^8$ MC cycles following an equilibration period of $2\times 10^7$ MC cycles.
Further details of the method can be found in Ref.~\onlinecite{polson2019polymer}.

\begin{figure}[!htb]
\begin{center}
\vspace*{0.2in}
\includegraphics[width=0.48\textwidth]{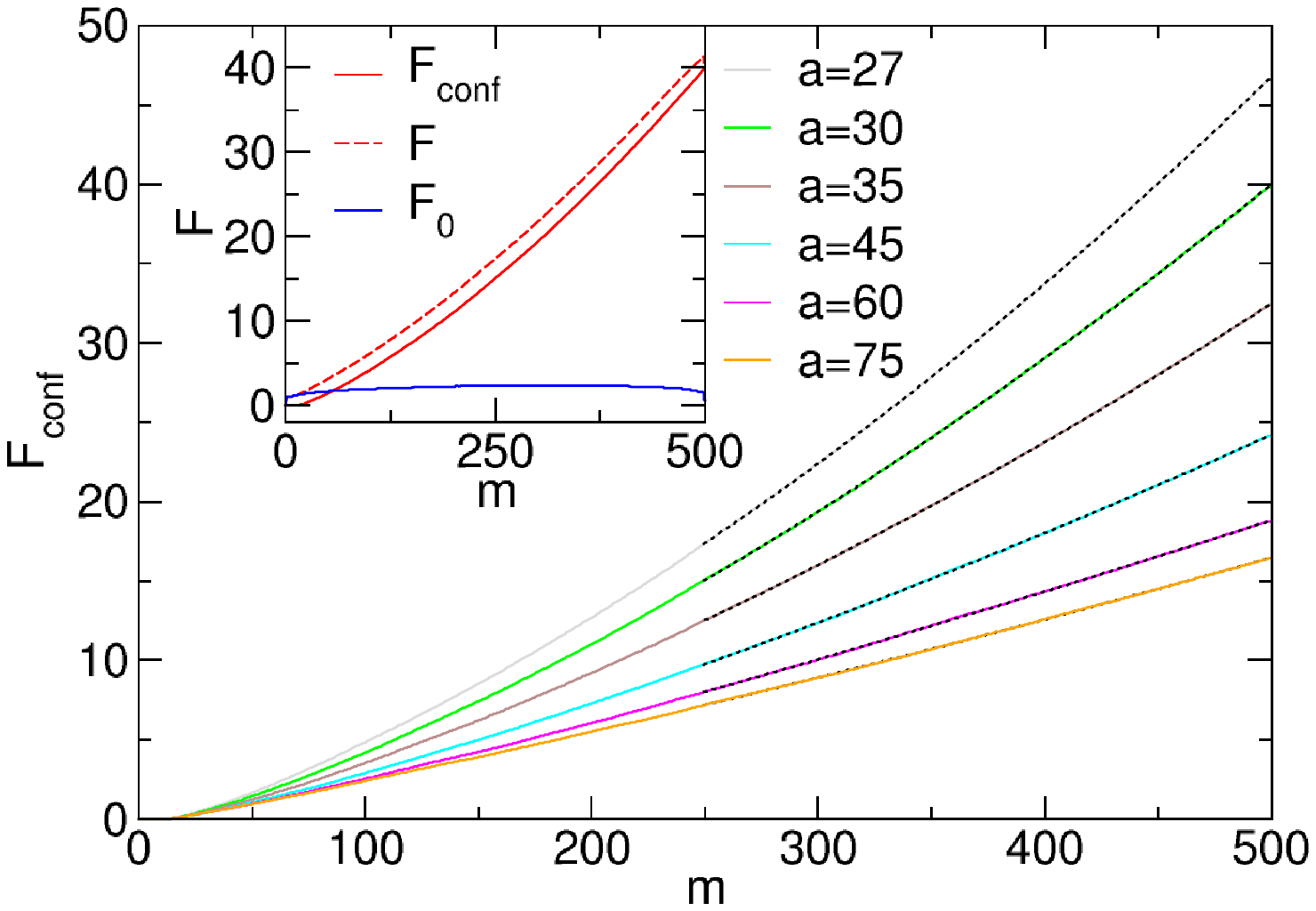}
\end{center}
\caption{Confinement free energy of a polymer chain of length $N=500$ with
bending rigidity $\kappa=5$. The chain is confined to a rectangular cavity
of dimensions $a\times a\times d$. Results are shown for $d=20$ and various
values of $a$. The inset illustrates the evaluation of $F_{\rm conf}$ as
the difference between $F_1(m)$ and $F_0(m)$, as described in the text.  }
\label{fig:Ftrans}
\end{figure}

Figure~\ref{fig:Ftrans} shows $F_{\rm conf}$ vs $m$ for a chain translocating into
a rectangular box with various values of the side length, $a$. The dotted black lines
overlaid on the curves show fits to the function $F=c_0 + c_1 m + Bm^2$, which was
carried out over the range of $m=250-500$. The inset shows the functions $F_1(m)$
and $F_{\rm conf}(m)$ for $a=30$, as well as the translocation free energy function
$F_0(m)$ in the case of no confinement on either side of the barrier.

%

\end{document}